\documentclass[sigconf]{acmart}

\AtBeginDocument{%
  \providecommand\BibTeX{{%
    \normalfont B\kern-0.5em{\scshape i\kern-0.25em b}\kern-0.8em\TeX}}}

\begin{document}

\title{Vibrotactile Feedback for Vertical 2D Space Exploration}

\author{Lancelot Dupont}
\email{lancelot.dupont@grenoble-inp.org}
\affiliation{%
\institution{Grenoble INP-Phelma}
\country{France\\}
  \institution{Singapore University of Technology and Design (SUTD)}
  \country{Singapore}
}

\author{Christophe Jouffrais}
\email{Christophe.Jouffrais@irit.fr}
\affiliation{%
  \institution{CNRS, IPAL}
  \country{Singapore\\}
  \institution{CNRS, IRIT}
  \country{Toulouse, France}
}

\author{Simon T. Perrault}
\email{perrault.simon@gmail.com}
\affiliation{%
  \institution{Singapore University of Technology and Design (SUTD)}
  \country{Singapore}
}

\renewcommand{\shortauthors}{Anonymous et al.}

\begin{abstract}
Visually impaired people encounter many challenges in their everyday life, especially when it comes to navigating and representing space. The issue of shopping is addressed mostly on the level of navigation and product detection, but conveying clues about the object position to the user is rarely implemented. This work presents a prototype of vibrotactile wristband using spatiotemporal patterns to help visually impaired users reach an object in the 2D plane in front of them. A pilot study on twelve blindfolded sighted subjects showed that discretizing space in a seven by seven targets matrix and conveying clues with a discrete pattern on the vertical axis and a continuous pattern on the horizontal axis is an intuitive and effective design.
\end{abstract}




\maketitle

\section{Introduction}
Visual impairment is a major obstacle to a person's autonomy in everyday life, especially in relation to space. Many works have been carried out and are underway to help navigation in space (\cite{Aggravi2016,Gharani2017,Johnson2006}) but the problem of close space seems to be less considered.
A task like shopping however requires both aspects: first the user needs to navigate through the shop and then precisely locate a specific object. That kind of task is perceived as particularly difficult by people whose eyesight is severely impaired \cite{Lamoureux2004}.

According to Kostyra et al~\cite{Kostyra2017}, around a third of visually impaired participants indicated that they run their basic necessities shopping online. The majority travel to the store, and 72\% said they do it alone. But the poor availability of sellers and lack of braille information on packaging makes this task frustrating and time consuming.
Locating one object precisely among others within a radius can be facilitated by the use of assistive devices, but few solutions exist to overcome these problems. The use of text-to-speech applications allows the labels to be read, however this does not make the search for a product faster.
Some devices already offer guidance through the shelves and the transmission of product information to the customer via audio feedback, for example BlindShopping~\cite{Lopez2011}. But guidance when catching the object is only rarely finalized, and when it is, audio is usually used, which masks the ambient noises necessary for visually impaired people to understand their environment.

In this paper, we propose a wearable device using vibrotactile feedback that provides spatial information to precisely locate an object close to the user in a reliable, discreet, rapid manner that does not mask the ambient sound cues. In our study, we considered two main factors. First, the division of space into discrete circular positions, hereinafter called \emph{Target Density}. Second, the methods of encoding the information transmitted along the vertical axis and along the horizontal axis, designated below by "combination of vibration codes" or \emph{Encoding}. Target position can be encoded in either a discrete or continuous manner.

Our results suggest that the best working encoding is a hybrid combination, relying on discrete encoding on the vertical axis, and continuous encoding on the horizontal one. The contribution of this paper is two-fold: (1) a wearable device with four vibration motors that can encode the position of objects in space and (2) the results of our user study in which we compare three \emph{Target Densities} and four \emph{Encodings}.

\section{Related Work}

\subsection{Challenges for Visually Impaired People}
A study by Papadopoulos et al~\cite{Papadopoulos2011} about the concerns of visually impaired children and their parents showed that the area most affected by their disability concerns life skills compared to communication skills and social skills. It has also shown that navigation capabilities are an indicator of performance and lag in development in this area. In addition, autonomous navigation would allow children to participate in more social activities and improve their social skills. For these reasons, mobility has been at the center of research and development for assistive devices in recent decades. Simple actions for a sighted person such as keeping a direction while walking present a strong cognitive load in the absence of visual feedback.

\subsection{Vibrotactile Guidance}
Navigation in space is a well addressed problem by researchers working on visual impairment. Many navigation aids based on a vibrotactile feedback have been developed, which suggests a good efficiency of this method. Kammoun et al~\cite{Kammoun2012} use two vibrating bracelets to help a blind person to maintain a direction while walking. Results were encouraging despite some confusion due to the vibration encodings.
Lim et al~\cite{Lim2015} offers another example of guidance based purely on haptic feedback. This article showed that it was possible to guide a sighted person in a shopping center only by means of vibration indices via a smart watch and a mobile phone.
Dobbelstein et al~\cite{Dobbelstein2016} have sought to accomplish the same goal by transmitting less information: only the general direction of the destination was communicated through a vibrotactile wristband while the subject was walking in the streets. The study showed a reduced cognitive load for the user, at the cost of navigation problems, e.g. pedestrian crossings without an audible signal or markings on the ground.
Aggravi et al~\cite{Aggravi2016} have developed a guidance system for blind skiers. The principle is that the person guiding the skier can send vibrotactile "left" or "right" instructions of different lengths using buttons in the handles of his poles.

\subsection{Locating an Object}
The problem of this research is to design a device helping the visually impaired in an object search task, e.g. shopping in a supermarket. Surveys have shown that visually impaired people regard shopping centers as the most difficult environments to navigate in~\cite{Passini1988}. In addition, shopping would be a major problem~\cite{Lamoureux2004} because it requires both information about nearby objects (e.g. reaching the desired object in a department) and the environment as a whole (e.g. finding a particular department in the store).
This type of task requires an independent and secure mobility of the subject, a problem commonly explored in research today as it was presented previously. The device presented in this article deals with the second task, reaching the object of your choice once you have arrived at its location.
People with tunnel vision have a clear field of vision but reduced to less than 10 degrees, located in the center of the retina, which deprives them of peripheral vision and therefore makes it difficult and time-consuming to find specific points of interest in a visual scene.
The work of Appert et al~\cite{Appert2015} studied techniques of vibrotactile interactions communicating the relative position of a target in the visual reference of the user. Their results show that encoding the distance using discrete vibrations in a Cartesian coordinates system is preferable to encoding with polar coordinates. Our work builds from Appert et al.'s and shows that for best results, an hybrid encoding using both discrete and continuous vibration in a cartesian frame of reference leads to the best results.

\section{Proposed Solution}
To help visually impaired users locate an object in a vertical 2D plane in front of them, we designed a wearable device that can be used to convey vibrotactile feedback to encode the position of an object.

\subsection{Hardware}
\begin{figure}
    \centering
     \includegraphics[width=0.8\columnwidth]{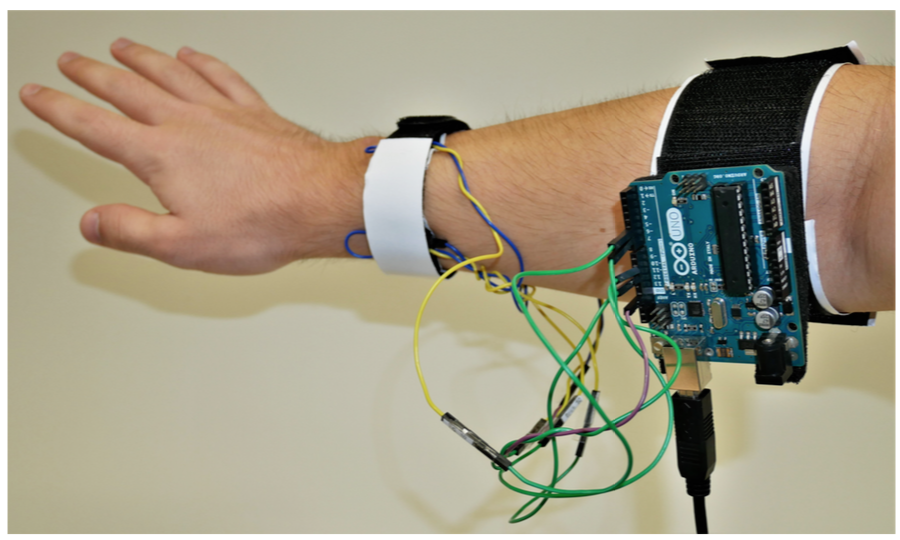}
    \caption{Hardware prototype used in the experiment. The wires going to the wristband are connected to the top and left vibration motors respectively.}~\label{fig:apparatus}
\end{figure}

We designed a wristband using Velcro, which allows us to adjust the device size as well as the locations of the vibration motors. We used four coin-type vibration motors with a diameter of 10 mm (Precision Microdrives, 310-103) operating at 130 Hz.
The four motors are distributed equally around the user's wrist, on Top, Left, Right and Bottom. Figure~\ref{fig:apparatus} shows the wristband.

The wristband is connected to an Arduino Uno micro-controller, which receives orders from the experimental software located on the experimenter's computer. The Arduino Uno is also connected to the laptop using a USB cable.
We initially considered using a wireless connection for the Arduino micro-controller but the connection was not always stable, thus we reverted to a wired solution (Figure~\ref{fig:apparatus}).

The experimental software handles the workflow for the experiment, and displays a view of the targets, allowing the experimenter to perform the target selection with a mouse. It also sends order to the Arduino to encode specific positions.

\subsection{Vibrotactile Encoding}
\begin{figure*}[t]
    \centering
     \includegraphics[width=1.4\columnwidth]{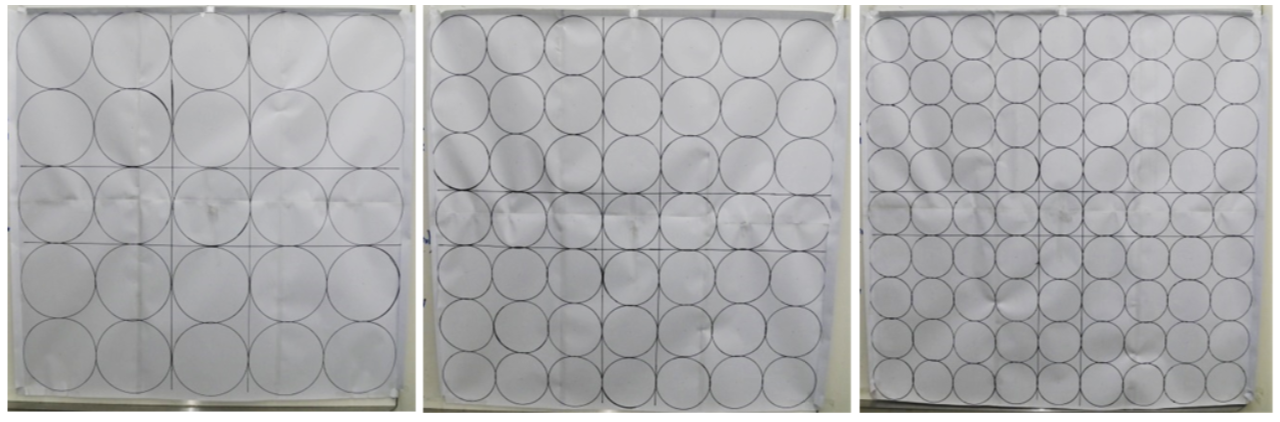}
    \caption{Target densities used in the experiment: $5 \times 5$, $7\times 7$, $9 \times 9$.}~\label{fig:targets}
\end{figure*}

We encode the position of an object in a 2D frame using X and Y coordinate.
The center of the 2D plane is located at the level of the user's sternum. Every target position is encoded from the origin of the frame and the frame has a dimension of $80 \times 80$ cm (Figure~\ref{fig:targets}).
Each of the four motors codes a different direction: Up, Left, Right, Down.
Following Appert et al~\cite{Appert2015}, we encoded the coordinates using either a discrete or continuous method on both axis.
For the discrete encoding, we used different number of pulses, where a low number of pulses represents a target closer to the origin. In that encoding, pulses last for exactly 200 milliseconds, with a 120 milliseconds pause between them.
For the continuous encoding, we used a varying duration of the pulses, with a shorter pulse showing a position closer to the origin of the frame, with a duration of vibration between 200 and 1300 milliseconds.

The encoding is transmitted sequentially, with the vertical encoding sent first, and the horizontal encoding sent afterwards, with a pause of 800 milliseconds between them. We avoided sending the encoding simultaneously following guidelines from Carcedo et al~\cite{Carcedo2016}.
In our experiment, we used an odd number of targets on both axis, resulting in some targets having a X or Y coordinate of 0 (e.g. the origin). For these special cases, the system will send a brief 120 ms pulse on the two motors encoding the axis with 40 ms in between. The origin point is thus coded with a 120 ms pulse on the top motor, followed by a 40 ms pause, 120 ms pulse on the bottom motor, 800 ms pause, 120 ms pulse on the left motor, 40 ms pause, 120 ms pulse  on the right motor.
The intensity of the vibration is always the same (100\% PWM, i.e. the maximum intensity available).

\section{Experiment}
We ran an experiment to understand which encoding allows blindfolded users to be the most accurate for different target densities.

\subsection{Participants}
Twelve participants (9 female, 11 right-handed), aged 19 to 33 ($M=22.4$, $SD=3.6$) were recruited from the university community and were given the equivalent of 4 euros for the participation. For this study we used blindfolded participants as their behavior and spatial acuity is similar to a late-blind user.

\subsection{Task and Stimuli}
At the start of each trial, blindfolded participants were instructed to put their dominant hand on the origin of the frame, which had a tactile mark to make it easy to find. The software would then send the encoding for the target they had to reach. The borders of the frame also had tactile marks, making it easier for participants to find the extremities.
Participants would then move their hand towards the target and notify the experimenter when they thought the position of their hand was accurate, by enumerating the expected position of the target (e.g. "two up, one left").
This allowed the experimenter to classify potential errors as either an incorrect interpretation of the encoding, or simply a positional error (i.e. the hand not reaching the exact position). Participants were allowed to receive the encoding a second time.
As soon as the participant felt confident with their choice, the experimenter would stop the timer on the experimental software by clicking on the selected target on the experimental software.

\subsection{Procedure}
At the beginning of the experiment, participants answered a questionnaire on their demographics. The experimenter would thus explain the purpose of the study, and help the participant put on the wristband, and adjust the location of the motors. The experimenter would then send a pulse on each four motors and ask the participants to correctly identify the position of each motor in space.
After that, the participant would be blindfolded and put in front of the frame, with the position of the frame adjusted to the sternum. Participants were placed at a distance that would allow them to easily reach each four corners of the frame with their arm in a comfortable position.

The participants were then trained on each encoding scheme, and would thus proceed with each three target densities. Within each target density all 4 vibrotactile encodings were tested. There was a short 2 seconds break between each trial. Each condition comprised one block, with 9 trials per condition. The 9 targets of each block were randomly selected from the 9 areas following areas: top left, top central column, top right, central row left, origin, central row right, bottom left, bottom central column, bottom right.

\subsection{Design}
We used a $3\times4$ within subject design with two indepent variables: \emph{Target Density} \{ $5 \times 5$, $7 \times 7$, $9 \times 9$ \} and \emph{Encoding} \{ Vertical-Discrete/Horizontal-Discrete (VDHD), Vertical-Discrete/Horizontal-Continuous (VDHC), Vertical-Continuous/Horizontal-Discrete (VCHD), Vertical-Continous/Horizontal-Continous (VCHC) \} . The order of presentation of both independent variable is counterbalanced using a Latin Square.
The experiment is divided into 12 blocks (1 per condition), with 9 trials per block.

We measured the time to acquire a target as well as accuracy as dependent variable. A trial was considered successful if the index finger of the participant was within the circle representing the target. Participants were encouraged to take breaks between blocks and completed the experiment in approximately 60 minutes.
Our overall design is as follows: 12 participants $\times$ 3 target densities $\times$ 4 encodings $\times$ 1 block per condition $\times$ 9 trials = 1296 trials.

\subsection{Results}
We used a two-way ANOVA with repeated measures on all factors to get our main effect for time, and a Friedman's test for accuracy. For post-hoc comparisons we used pairwise t-tests with Bonferroni correction for time, or pairwise Wilcoxon tests with Bonferroni-Holm correction for accuracy.

\subsubsection{Time}
\begin{figure}[h]
    \centering
     \includegraphics[width=1\columnwidth]{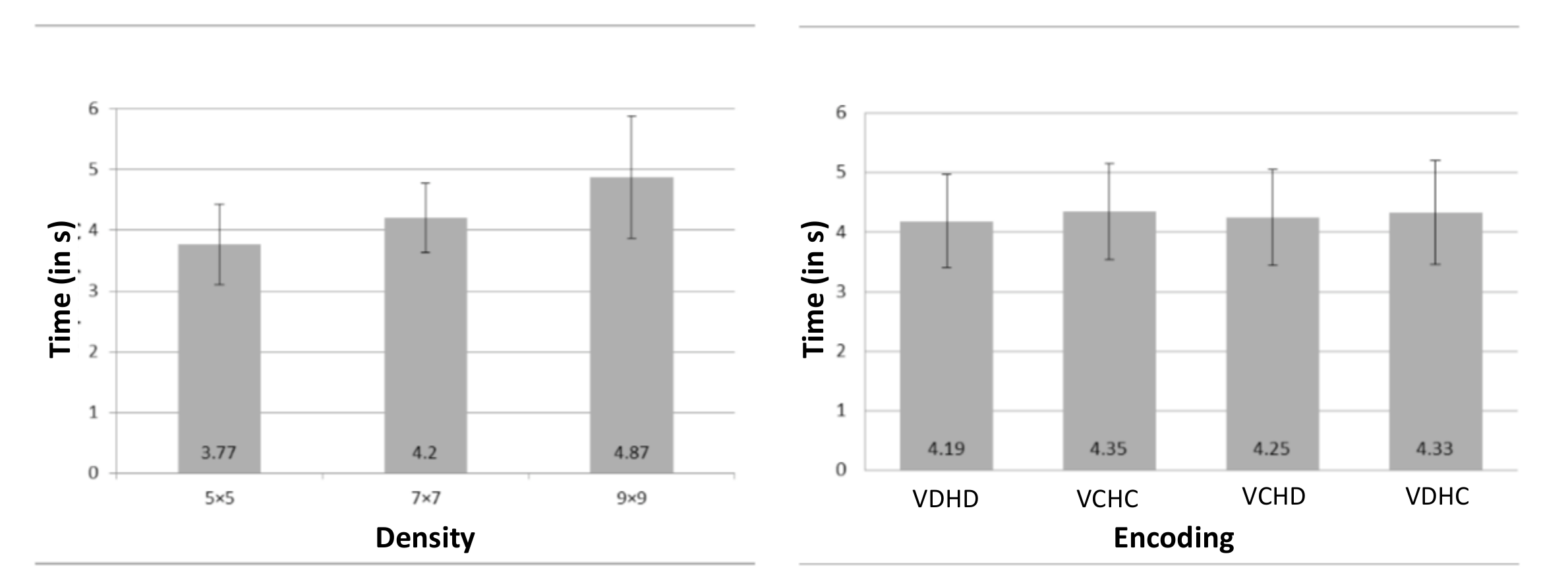}
    \caption{Average trial time for each \emph{Target Density} and \emph{Encoding} condition. Error bars show .95 confidence intervals.}~\label{fig:time}
\end{figure}

Our time data did not follow a normal distribution, and we thus used a logarithmic transformation to perform our ANOVA. After transforming the data, we found a significant main effect of \emph{Target Density} ($F_{2,22}=6.53$, $p<.01$) on time. Post-hoc analysis showed that our participants were significantly faster in the $5 \times 5$ density ($M=3.77s$) compared to $7 \times 7$ ($M=4.2s$) and $9 \times 9$ ($M=4.87s$, both $p<.01$). We also found a significant difference between the $7 \times 7$ and $9 \times 9$ densities ($p<.001$).
We did not find any significant main effect of the \emph{Encoding} ($p=.96$) with average time performance ranging from 4.19s (VDHD) to 4.33s (HCVD) or any interaction ($p=.97$). Time performance is summarized in Figure~\ref{fig:time}.

\subsubsection{Accuracy}
\begin{figure}[t]
    \centering
     \includegraphics[width=1\columnwidth]{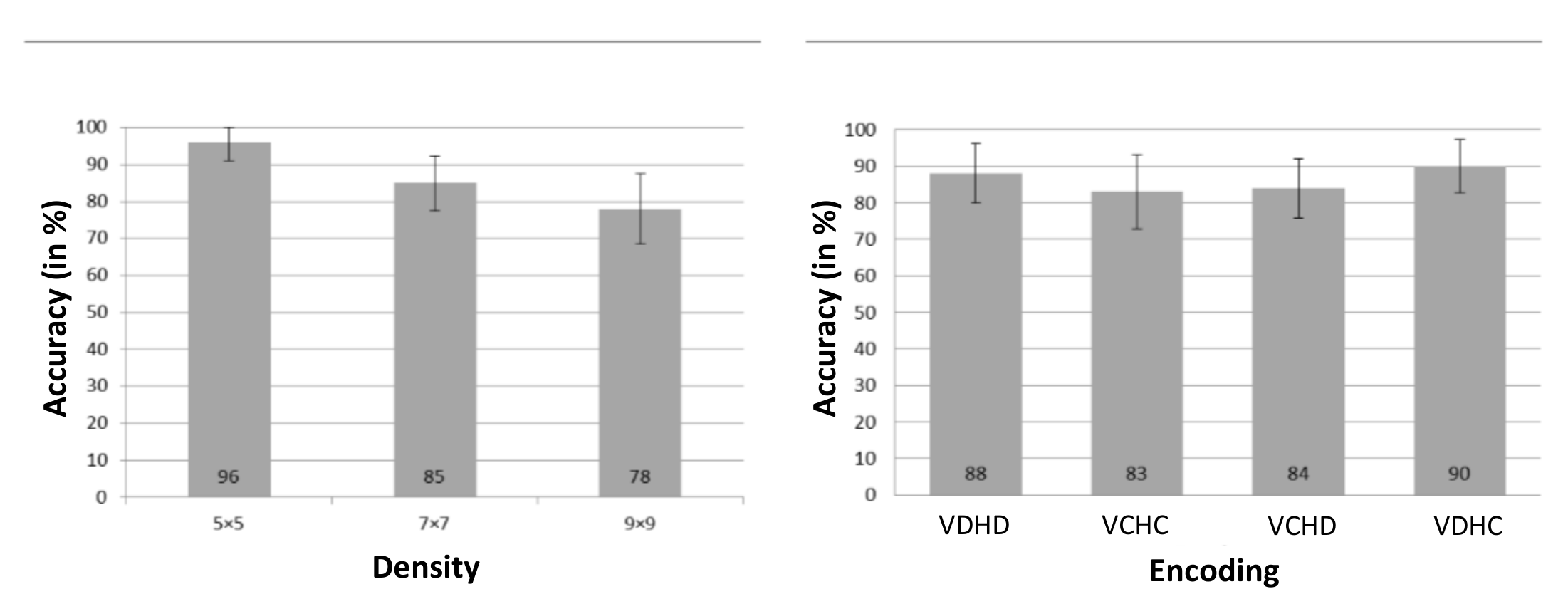}
    \caption{Average accuracy for each \emph{Target Density} and \emph{Encoding} condition. Error bars show .95 confidence intervals.}~\label{fig:accuracy}
\end{figure}

Our participants were overall quite accurate with an average accuracy of 87\%. Among the 13\% error rates, 6\% is due to an incorrect understanding of the encoding (e.g. confusion top vs. bottom or left vs. right) and 7\% due to positional errors.
We found a significant main effect of \emph{Target Density} on accuracy ($\chi^{2}(2)=9.04$, $p<.01$). Participants were significantly more accurate in the $5 \times 5$ density ($M=96\%$) compared to $7 \times 7$ ($M=85\%$, $p<.05$) and $9 \times 9$ ($M=78\%$, $p<.01$).
We also found a significant main effect of \emph{Encoding} on accuracy ($\chi^{2}(3)=10.23$, $p<.05$). Post-hoc analysis reveals that participants tended to be more accurate in the VDHC condition ($M=90.4\%$) as compared to VCHC ($M=83.6\%$, $p<.05$) and VCHD ($M=84.3\%$, $p<.05$).
Accuracy performance is illustrated in Figure~\ref{fig:accuracy}.

\section{Discussion}
Overall, our results suggest that our participants were able to accurately locate targets in space with a low error rate. The highest density suggests that participants were able to locate a circular target of 9 cm radius nearly 4 times out of 5 (78\%) with limited training.
The accuracy is even better in the $5 \times 5$ (96\%) and $7 \times 7$ (85\%), which represent a target radius of 16 cm and 11.5 cm respectively. None of the encoding we designed seemed to allow participants to identify a target faster, however, the Vertical-Discrete/Horizontal-Continuous encoding led to significantly higher accuracy overall.

Discretizing vertical space can be convenient in any scenario when an object is located on a shelf, as there is a natural mapping between a number of pulse and a "shelf number" which discretizes the physical space as well.
For that reason, the VCHD seems like the best choice for real life usage.
Our participants reported that they generally preferred techniques where the information was discrete as opposed to continuous, with their favorite technique being VDHD, as counting pulses is easier than mapping a time duration to a coordinate. This result is in line with Carcedo et al~\cite{Carcedo2016}.

\section{Limitations and Future Work}
This experiment is a first step towards the conception of a new generation of assistive devices that can accurately help visually impaired users to locate objects in a 2D frame in front of them.
In this experiment, we used blindfolded users, which can be compared to late-blind users, but future work should use visually impaired users to confirm our results.
In addition, we used a physical apparatus with tactile marks on the edges and center of our frame, which may be hard to replicate in real life.
Finally, this study only investigates the best type of vibrotactile feedback to help guide the users but does not provide a recovery mechanism whenever a wrong target is selected and also does not offer any input solution to recognize specific objects.\\
\\
We do believe that our results are encouraging and provide an interesting encoding system that users can quickly learn to get a better understanding of the space in front of them.

\section{Conclusion}
In this paper, we developed a wristband prototype embedding four vibration sensors. In a controlled experiment, we compared multiple vibrotactile encodings to locate targets in a $80 \times 80$ cm 2D frame in front of them. Our results suggest that blindfolded participants are able to accurately locate object with a 11.5 to 16 cm radius in that space with more than 85\% accuracy. More specifically, we showed that the most accurate encoding uses a discrete encoding for vertical axis and continuous encoding for the horizontal axis, which maps very well with scenarios where the user is trying to find objects on shelves.
\bibliographystyle{ACM-Reference-Format}
\bibliography{biblio}


\begin{thebibliography}{13}


\ifx \showCODEN    \undefined \def \showCODEN     #1{\unskip}     \fi
\ifx \showDOI      \undefined \def \showDOI       #1{#1}\fi
\ifx \showISBNx    \undefined \def \showISBNx     #1{\unskip}     \fi
\ifx \showISBNxiii \undefined \def \showISBNxiii  #1{\unskip}     \fi
\ifx \showISSN     \undefined \def \showISSN      #1{\unskip}     \fi
\ifx \showLCCN     \undefined \def \showLCCN      #1{\unskip}     \fi
\ifx \shownote     \undefined \def \shownote      #1{#1}          \fi
\ifx \showarticletitle \undefined \def \showarticletitle #1{#1}   \fi
\ifx \showURL      \undefined \def \showURL       {\relax}        \fi
\providecommand\bibfield[2]{#2}
\providecommand\bibinfo[2]{#2}
\providecommand\natexlab[1]{#1}
\providecommand\showeprint[2][]{arXiv:#2}

\bibitem[\protect\citeauthoryear{Aggravi, Salvietti, and Prattichizzo}{Aggravi
  et~al\mbox{.}}{2016}]%
        {Aggravi2016}
\bibfield{author}{\bibinfo{person}{Marco Aggravi}, \bibinfo{person}{Gionata
  Salvietti}, {and} \bibinfo{person}{Domenico Prattichizzo}.}
  \bibinfo{year}{2016}\natexlab{}.
\newblock \showarticletitle{Haptic Assistive Bracelets for Blind Skier
  Guidance}. In \bibinfo{booktitle}{\emph{Proceedings of the 7th Augmented
  Human International Conference 2016}} (Geneva, Switzerland)
  \emph{(\bibinfo{series}{AH ’16})}. \bibinfo{publisher}{Association for
  Computing Machinery}, \bibinfo{address}{New York, NY, USA}, Article
  \bibinfo{articleno}{Article 25}, \bibinfo{numpages}{4}~pages.
\newblock
\showISBNx{9781450336802}
\urldef\tempurl%
\url{https://doi.org/10.1145/2875194.2875249}
\showDOI{\tempurl}


\bibitem[\protect\citeauthoryear{Appert, Camors, Durand, and Jouffrais}{Appert
  et~al\mbox{.}}{2015}]%
        {Appert2015}
\bibfield{author}{\bibinfo{person}{Damien Appert}, \bibinfo{person}{Damien
  Camors}, \bibinfo{person}{Jean-Baptiste Durand}, {and}
  \bibinfo{person}{Christophe Jouffrais}.} \bibinfo{year}{2015}\natexlab{}.
\newblock \showarticletitle{Tactile Cues for Improving Target Localization in
  Subjects with Tunnel Vision}. In \bibinfo{booktitle}{\emph{Proceedings of the
  27th Conference on l’Interaction Homme-Machine}} (Toulouse, France)
  \emph{(\bibinfo{series}{IHM ’15})}. \bibinfo{publisher}{Association for
  Computing Machinery}, \bibinfo{address}{New York, NY, USA}, Article
  \bibinfo{articleno}{Article 6}, \bibinfo{numpages}{10}~pages.
\newblock
\showISBNx{9781450338448}
\urldef\tempurl%
\url{https://doi.org/10.1145/2820619.2820625}
\showDOI{\tempurl}


\bibitem[\protect\citeauthoryear{Carcedo, Chua, Perrault, Wozniak, Joshi,
  Obaid, Fjeld, and Zhao}{Carcedo et~al\mbox{.}}{2016}]%
        {Carcedo2016}
\bibfield{author}{\bibinfo{person}{Marta~G. Carcedo}, \bibinfo{person}{Soon~Hau
  Chua}, \bibinfo{person}{Simon Perrault}, \bibinfo{person}{Pawe\l{} Wozniak},
  \bibinfo{person}{Raj Joshi}, \bibinfo{person}{Mohammad Obaid},
  \bibinfo{person}{Morten Fjeld}, {and} \bibinfo{person}{Shengdong Zhao}.}
  \bibinfo{year}{2016}\natexlab{}.
\newblock \showarticletitle{HaptiColor: Interpolating Color Information as
  Haptic Feedback to Assist the Colorblind}. In
  \bibinfo{booktitle}{\emph{Proceedings of the 2016 CHI Conference on Human
  Factors in Computing Systems}} (San Jose, California, USA)
  \emph{(\bibinfo{series}{CHI ’16})}. \bibinfo{publisher}{Association for
  Computing Machinery}, \bibinfo{address}{New York, NY, USA},
  \bibinfo{pages}{3572–3583}.
\newblock
\showISBNx{9781450333627}
\urldef\tempurl%
\url{https://doi.org/10.1145/2858036.2858220}
\showDOI{\tempurl}


\bibitem[\protect\citeauthoryear{Dobbelstein, Henzler, and Rukzio}{Dobbelstein
  et~al\mbox{.}}{2016}]%
        {Dobbelstein2016}
\bibfield{author}{\bibinfo{person}{David Dobbelstein}, \bibinfo{person}{Philipp
  Henzler}, {and} \bibinfo{person}{Enrico Rukzio}.}
  \bibinfo{year}{2016}\natexlab{}.
\newblock \showarticletitle{Unconstrained Pedestrian Navigation Based on
  Vibro-Tactile Feedback around the Wristband of a Smartwatch}. In
  \bibinfo{booktitle}{\emph{Proceedings of the 2016 CHI Conference Extended
  Abstracts on Human Factors in Computing Systems}} (San Jose, California, USA)
  \emph{(\bibinfo{series}{CHI EA ’16})}. \bibinfo{publisher}{Association for
  Computing Machinery}, \bibinfo{address}{New York, NY, USA},
  \bibinfo{pages}{2439–2445}.
\newblock
\showISBNx{9781450340823}
\urldef\tempurl%
\url{https://doi.org/10.1145/2851581.2892292}
\showDOI{\tempurl}


\bibitem[\protect\citeauthoryear{Gharani and Karimi}{Gharani and
  Karimi}{2017}]%
        {Gharani2017}
\bibfield{author}{\bibinfo{person}{Pedram Gharani} {and}
  \bibinfo{person}{Hassan~A. Karimi}.} \bibinfo{year}{2017}\natexlab{}.
\newblock \showarticletitle{Context-aware obstacle detection for navigation by
  visually impaired}.
\newblock \bibinfo{journal}{\emph{Image and Vision Computing}}
  \bibinfo{volume}{64} (\bibinfo{year}{2017}), \bibinfo{pages}{103 -- 115}.
\newblock
\showISSN{0262-8856}
\urldef\tempurl%
\url{https://doi.org/10.1016/j.imavis.2017.06.002}
\showDOI{\tempurl}


\bibitem[\protect\citeauthoryear{{Johnson} and {Higgins}}{{Johnson} and
  {Higgins}}{2006}]%
        {Johnson2006}
\bibfield{author}{\bibinfo{person}{L.~A. {Johnson}} {and}
  \bibinfo{person}{C.~M. {Higgins}}.} \bibinfo{year}{2006}\natexlab{}.
\newblock \showarticletitle{A Navigation Aid for the Blind Using Tactile-Visual
  Sensory Substitution}. In \bibinfo{booktitle}{\emph{2006 International
  Conference of the IEEE Engineering in Medicine and Biology Society}}.
  \bibinfo{pages}{6289--6292}.
\newblock
\showISSN{1557-170X}
\urldef\tempurl%
\url{https://doi.org/10.1109/IEMBS.2006.259473}
\showDOI{\tempurl}


\bibitem[\protect\citeauthoryear{Kammoun, Jouffrais, Guerreiro, Nicolau, and
  Jorge}{Kammoun et~al\mbox{.}}{2012}]%
        {Kammoun2012}
\bibfield{author}{\bibinfo{person}{Slim Kammoun}, \bibinfo{person}{Christophe
  Jouffrais}, \bibinfo{person}{Tiago Guerreiro}, \bibinfo{person}{Hugo
  Nicolau}, {and} \bibinfo{person}{Joaquim Jorge}.}
  \bibinfo{year}{2012}\natexlab{}.
\newblock \showarticletitle{Guiding blind people with haptic feedback}.
\newblock \bibinfo{journal}{\emph{Frontiers in Accessibility for Pervasive
  Computing (Pervasive 2012)}}  \bibinfo{volume}{3} (\bibinfo{year}{2012}).
\newblock


\bibitem[\protect\citeauthoryear{Kostyra, Żakowska Biemans, Śniegocka, and
  Piotrowska}{Kostyra et~al\mbox{.}}{2017}]%
        {Kostyra2017}
\bibfield{author}{\bibinfo{person}{Eliza Kostyra}, \bibinfo{person}{Sylwia
  Żakowska Biemans}, \bibinfo{person}{Katarzyna Śniegocka}, {and}
  \bibinfo{person}{Anna Piotrowska}.} \bibinfo{year}{2017}\natexlab{}.
\newblock \showarticletitle{Food shopping, sensory determinants of food choice
  and meal preparation by visually impaired people. Obstacles and expectations
  in daily food experiences}.
\newblock \bibinfo{journal}{\emph{Appetite}}  \bibinfo{volume}{113}
  (\bibinfo{year}{2017}), \bibinfo{pages}{14 -- 22}.
\newblock
\showISSN{0195-6663}
\urldef\tempurl%
\url{https://doi.org/10.1016/j.appet.2017.02.008}
\showDOI{\tempurl}


\bibitem[\protect\citeauthoryear{Lamoureux, Hassell, and Keeffe}{Lamoureux
  et~al\mbox{.}}{2004}]%
        {Lamoureux2004}
\bibfield{author}{\bibinfo{person}{Ecosse~L Lamoureux},
  \bibinfo{person}{Jennifer~B Hassell}, {and} \bibinfo{person}{Jill~E Keeffe}.}
  \bibinfo{year}{2004}\natexlab{}.
\newblock \showarticletitle{The determinants of participation in activities of
  daily living in people with impaired vision}.
\newblock \bibinfo{journal}{\emph{American Journal of Ophthalmology}}
  \bibinfo{volume}{137}, \bibinfo{number}{2} (\bibinfo{year}{2004}),
  \bibinfo{pages}{265 -- 270}.
\newblock
\showISSN{0002-9394}
\urldef\tempurl%
\url{https://doi.org/10.1016/j.ajo.2003.08.003}
\showDOI{\tempurl}


\bibitem[\protect\citeauthoryear{Lim, Cho, Rhee, and Suh}{Lim
  et~al\mbox{.}}{2015}]%
        {Lim2015}
\bibfield{author}{\bibinfo{person}{Hyunchul Lim}, \bibinfo{person}{YoonKyong
  Cho}, \bibinfo{person}{Wonjong Rhee}, {and} \bibinfo{person}{Bongwon Suh}.}
  \bibinfo{year}{2015}\natexlab{}.
\newblock \showarticletitle{Vi-Bros: Tactile Feedback for Indoor Navigation
  with a Smartphone and a Smartwatch}. In \bibinfo{booktitle}{\emph{Proceedings
  of the 33rd Annual ACM Conference Extended Abstracts on Human Factors in
  Computing Systems}} (Seoul, Republic of Korea) \emph{(\bibinfo{series}{CHI EA
  ’15})}. \bibinfo{publisher}{Association for Computing Machinery},
  \bibinfo{address}{New York, NY, USA}, \bibinfo{pages}{2115–2120}.
\newblock
\showISBNx{9781450331463}
\urldef\tempurl%
\url{https://doi.org/10.1145/2702613.2732811}
\showDOI{\tempurl}


\bibitem[\protect\citeauthoryear{L{\'o}pez-de Ipi{\~{n}}a, Lorido, and
  L{\'o}pez}{L{\'o}pez-de Ipi{\~{n}}a et~al\mbox{.}}{2011}]%
        {Lopez2011}
\bibfield{author}{\bibinfo{person}{Diego L{\'o}pez-de Ipi{\~{n}}a},
  \bibinfo{person}{Tania Lorido}, {and} \bibinfo{person}{Unai L{\'o}pez}.}
  \bibinfo{year}{2011}\natexlab{}.
\newblock \showarticletitle{BlindShopping: Enabling Accessible Shopping for
  Visually Impaired People through Mobile Technologies}. In
  \bibinfo{booktitle}{\emph{Toward Useful Services for Elderly and People with
  Disabilities}}, \bibfield{editor}{\bibinfo{person}{Bessam Abdulrazak},
  \bibinfo{person}{Sylvain Giroux}, \bibinfo{person}{Bruno Bouchard},
  \bibinfo{person}{H{\'e}l{\`e}ne Pigot}, {and} \bibinfo{person}{Mounir
  Mokhtari}} (Eds.). \bibinfo{publisher}{Springer Berlin Heidelberg},
  \bibinfo{address}{Berlin, Heidelberg}, \bibinfo{pages}{266--270}.
\newblock
\showISBNx{978-3-642-21535-3}


\bibitem[\protect\citeauthoryear{Papadopoulos, Metsiou, and
  Agaliotis}{Papadopoulos et~al\mbox{.}}{2011}]%
        {Papadopoulos2011}
\bibfield{author}{\bibinfo{person}{Konstantinos Papadopoulos},
  \bibinfo{person}{Katerina Metsiou}, {and} \bibinfo{person}{Ioannis
  Agaliotis}.} \bibinfo{year}{2011}\natexlab{}.
\newblock \showarticletitle{Adaptive behavior of children and adolescents with
  visual impairments}.
\newblock \bibinfo{journal}{\emph{Research in Developmental Disabilities}}
  \bibinfo{volume}{32}, \bibinfo{number}{3} (\bibinfo{year}{2011}),
  \bibinfo{pages}{1086 -- 1096}.
\newblock
\showISSN{0891-4222}
\urldef\tempurl%
\url{https://doi.org/10.1016/j.ridd.2011.01.021}
\showDOI{\tempurl}


\bibitem[\protect\citeauthoryear{Passini and Proulx}{Passini and
  Proulx}{1988}]%
        {Passini1988}
\bibfield{author}{\bibinfo{person}{Romedi Passini} {and}
  \bibinfo{person}{Guyltne Proulx}.} \bibinfo{year}{1988}\natexlab{}.
\newblock \showarticletitle{Wayfinding without Vision: An Experiment with
  Congenitally Totally Blind People}.
\newblock \bibinfo{journal}{\emph{Environment and Behavior}}
  \bibinfo{volume}{20}, \bibinfo{number}{2} (\bibinfo{year}{1988}),
  \bibinfo{pages}{227--252}.
\newblock
\urldef\tempurl%
\url{https://doi.org/10.1177/0013916588202006}
\showDOI{\tempurl}
\showeprint{https://doi.org/10.1177/0013916588202006}


\end{thebibliography}


\end{document}